\documentclass[
twocolumn,
superscriptaddress,
amsmath,
amssymb,
aps,
pre,
]{revtex4-1}

\usepackage{graphicx}   
\usepackage{dcolumn}    
\usepackage{bm}         
\usepackage{color}      
\usepackage[extension=xxx]{hyperref}
\usepackage{physics}
\usepackage[bb=boondox]{mathalfa}

\begin{document}

\title{Heat rectification with a minimal model of two harmonic oscillators}


\author{M. A. Sim\'{o}n}
\email[]{miguelangel.simon@ehu.eus}
\affiliation{Departamento de Qu\'{i}mica-F\'{i}sica, Universidad del Pa\'{i}s Vasco, UPV- EHU - Bilbao, Spain}

\author{A. Ala\~{n}a}
\affiliation{Departamento de Qu\'{i}mica-F\'{i}sica, Universidad del Pa\'{i}s Vasco, UPV- EHU - Bilbao, Spain}

\author{M. Pons}
\affiliation{Departamento de F\'{i}sica Aplicada I, Universidad del Pa\'{i}s Vasco, UPV- EHU - Bilbao, Spain}

\author{A. Ruiz-Garc\'{i}a}
\affiliation{Departamento de F\'{i}sica, Universidad de La Laguna, La Laguna 38203, Spain}
\affiliation{IUdEA Instituto Universitario de Estudios Avanzados, Universidad de La Laguna, La Laguna 38203, Spain}

\author{J. G. Muga}
\email[]{jg.muga@ehu.es}
\affiliation{Departamento de Qu\'{i}mica-F\'{i}sica, Universidad del Pa\'{i}s Vasco, UPV- EHU - Bilbao, Spain}

\begin{abstract}
We study heat rectification in a minimalistic model composed of two masses subjected to on-site and coupling
linear forces in contact with effective Langevin baths induced by laser interactions.  Analytic expressions of the heat currents in the steady state are spelled out.  Asymmetric heat transport is found in this linear system if both the bath temperatures and the temperature dependent bath-system couplings
are also exchanged.
\end{abstract}

\maketitle


\section{Introduction \label{sec:Introduction}}
Heat rectification, firstly observed in 1936 by Starr \cite{Starr1936}, is the physical phenomenon, analogous to electrical current rectification in diodes, in which heat current through a device or medium is not symmetric with respect to the exchange of the baths at the boundaries. In the limiting case the device allows heat to propagate in one direction from the hot to the cold bath while it behaves as a thermal insulator in the opposite direction when the baths are exchanged.  In 2002 a paper by Terraneo \textit{et al.} \cite{Terraneo2002} demonstrated heat rectification numerically for a chain of nonlinear oscillators in contact with two thermal baths at different temperatures. Since then, there has been a growing interest in heat rectification  \cite{Pereira2019,Roberts2011,Li2012,Ye2017,Wang2008,Wang2007,Casati2006,Joulain2016,Chang2006,Kobayashi2009,Leitner2013,Elzouka2017,Pons2017,Alexander2020}, and the field remains very active because of the potential applications in fundamental science and technology, and the
fact that none of the proposals so far appears to be efficient and robust for
practical purposes.

Much effort has been devoted to  understand the underlying physical mechanism responsible for  rectification \cite{Pereira2019}.
In early times some kind of anharmonicity,  i.e. non-linear forces, in the substrate potential or in the particle-particle interactions, was identified as a fundamental requisite for rectification \cite{Li2012,Li2008,Hu2006,Zeng2008,Katz2016,Benenti2016}. This non-harmonic behavior leads to a temperature dependence of the phonon bands. The match/mismatch of the phonon bands (power spectra) governs the heat transport in the chain, allowing it when the bands match or obstructing it if they mismatch \cite{Terraneo2002,Li2004}. However, a work by Pereira \textit{et al.} \cite{Pereira2017} showed that rectification can also be found in effective harmonic systems if two requirements are met: some kind of structural asymmetry, and features that depend on the temperature so they change as the baths are inverted. Indeed,  in this article we demonstrate rectification in a minimalistic model of two harmonic oscillators where the coupling to the baths depends on the temperature.
This will be justified with a particular physical set up with trapped ions and lasers.

The article is organized as follows. In Section \ref{sec:Physical_Model}
we describe the physical model and its dynamical equations. In Section \ref{sec:covMatrix} we describe the dynamics of the system in terms of a covariance matrix. We also derive a set of algebraic equations that gives as solution the covariance matrix in the steady state. In Section \ref{sec:solutions} we solve the covariance matrix equations and find analytical expressions for the steady-state temperatures of the masses and heat currents. In Section \ref{sec:TrappedIonSetUp} we relate the parameters of our model to those in a physical set-up of Doppler cooled trapped ions. In Section \ref{sec:lookingForR} we make a parameter sweep looking for configurations which yield high rectification. We also study the power spectra of the oscillators, which confirm the match/mismatch patterns in cases where there is rectification. In Section \ref{sec:Conclusions} we summarize our results and present our conclusions.

\begin{figure}
  \includegraphics[width=1.1\linewidth]{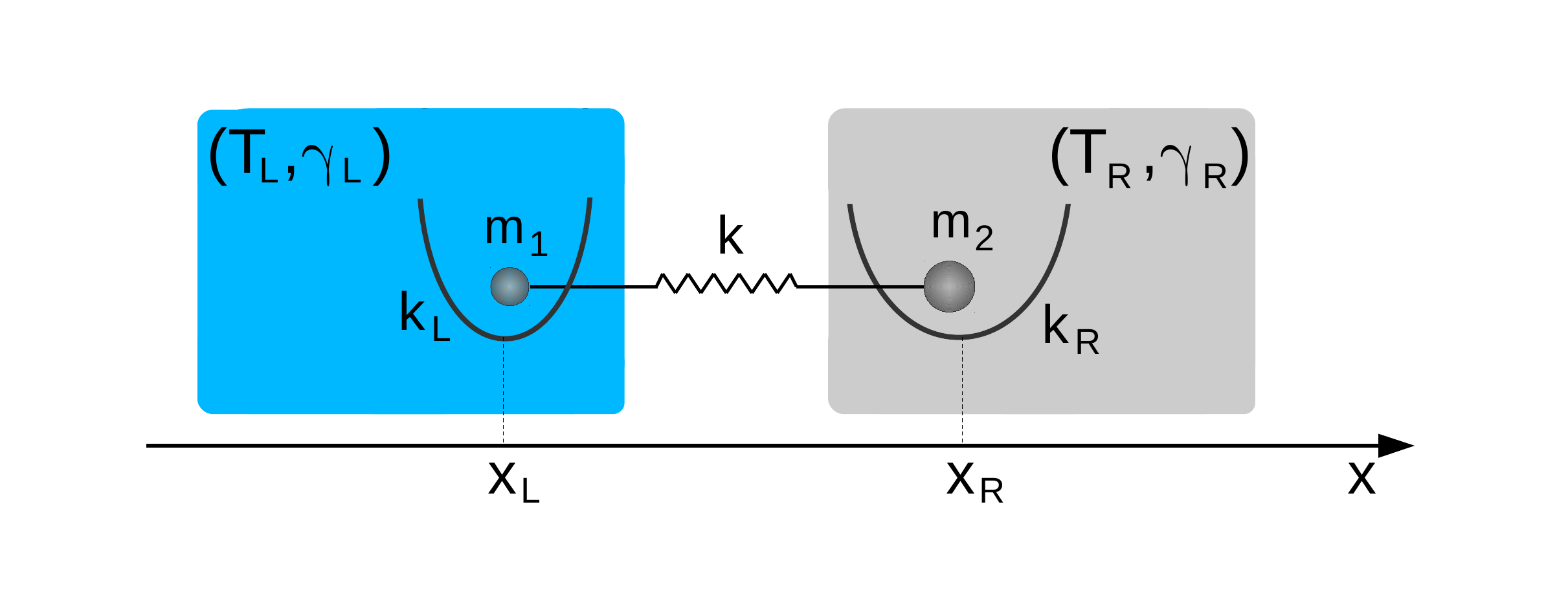}
  \caption{Diagram of the model described in Section \ref{sec:Physical_Model}. Two ions coupled to each other through a spring constant $k$. Each ion is harmonically trapped and connected to a bath characterized by its temperature $T_i$ and its friction coefficient $\gamma_i$. }
  \label{fig:model_diagram}
\end{figure}

\section{Physical Model \label{sec:Physical_Model}}

The physical model consists of two masses $m_1$ and $m_2$ coupled to each other by a harmonic interaction with spring constant $k$ and natural length $x_e$. Each of the masses $m_1$ and $m_2$ are confined by a harmonic potential with spring constants $k_L$, $k_R$ and equilibrium positions $x_L$, $x_R$ respectively (see Fig. \ref{fig:model_diagram}). The Hamiltonian describing this model is
\begin{equation}
  H = \frac{p_1^2}{2m_1} + \frac{p_2^2}{2m_2} + V(x_1,x_2),
  \label{eq:HamiltonianOriginalCordinates}
\end{equation}
with $V(x_1,x_2)=\frac{k}{2}\left( x_1 - x_2 - x_e \right)^2 + \frac{k_L}{2}\left( x_1 - x_L \right)^2 + \frac{k_R}{2}\left( x_2 - x_R \right)^2$,  where $\{x_i,p_i\}_{i=1,2}$ are the position and momentum of each mass. Switching from the original coordinates $x_i$ to displacements with respect to the equilibrium positions of the system $q_i = x_i - x_i^{eq}$, where $x_i^{eq}$ are the solutions to $\partial_{x_i}V(x_1,x_2)=0$, the Hamiltonian can be written as
\begin{align}
  H &= \frac{p_1^2}{2m_1} + \frac{p_2^2}{2m_2} + \frac{k+k_L}{2}q_1^2\nonumber\\ &+ \frac{k+k_R}{2}q_2^2 - k q_1 q_2 + V(x_1^{eq},x_2^{eq}).
  \label{eq:Hamiltonian}
\end{align}
This has the form of  the Hamiltonian of a system around a stable equilibrium point
\begin{equation}
  H = \frac{1}{2} \overrightarrow{p}^\mathsf{T}\mathbb{M}^{-1}\overrightarrow{p} + \frac{1}{2} \overrightarrow{q}^\mathsf{T}\mathbb{K}\overrightarrow{q},
\label{generic}
\end{equation}
where $\overrightarrow{q} = \left(q_1,q_2\right)^\mathsf{T}$, $\overrightarrow{p} = \left(p_1,p_2\right)^\mathsf{T}$, $\mathbb{M} = diag(m_1,m_2)$ is the mass matrix of the system and $\mathbb{K}$ is the Hessian matrix of the potential at the equilibrium point, i.e., $\mathbb{K}_{ij} = \partial^2_{x_i,x_j}V(\overrightarrow{x})\Big|_{\overrightarrow{x} = \overrightarrow{x}^{eq}}$. In this model  $\mathbb{K}_{11} = k + k_L$, $\mathbb{K}_{22} = k + k_R$ and $\mathbb{K}_{12} = \mathbb{K}_{21} = -k$.
We shall see later that
the generic form (\ref{generic}) can be adapted to different physical settings, in particular to
two ions in individual traps, or to two ions in a common trap.

The  masses are in contact with Langevin baths, which will be denoted as $L$ (for left) and $R$ (for right), at temperatures $T_{L}$ and $T_R$ for  the mass $m_1$ and $m_2$ respectively (see Fig. \ref{fig:model_diagram}). The equations of motion of the system, taking into account the Hamiltonian and the Langevin baths are
\begin{align}
  \dot{q}_1 &= \frac{p_1}{m_1},\nonumber
  \\
  \dot{q}_2 &= \frac{p_2}{m_2},\nonumber
  \\
  \dot{p}_1 &= -(k+k_L)q_1 + k q_2 -\frac{\gamma_L}{m_1} p_1 + \xi_L(t),\nonumber
  \\
  \dot{p}_2 &= -(k+k_R)q_2 + k q_1 -\frac{\gamma_R}{m_2} p_2 + \xi_R(t),
\end{align}
where $\gamma_L$, $\gamma_R$ are the friction coefficients of the baths and $\xi_L(t)$, $\xi_R(t)$ are Gaussian white-noise-like forces. The Gaussian forces have zero mean ($\expval{ \xi_L(t) } = \expval{ \xi_R(t) } = 0 $) and satisfy the correlations $\expval{ \xi_L(t)\xi_R(t') } = 0$, $\expval{ \xi_L(t)\xi_L(t') } = 2D_L\delta(t-t')$, $\expval{ \xi_R(t)\xi_R(t') } = 2D_R\delta(t-t')$. $D_L$ and $D_R$ are the diffusion coefficients, which satisfy the fluctuation-dissipation theorem: $D_L = \gamma_L k_B T_L$, $D_R =\gamma_R k_B T_R$ ($k_B$ is the Boltzmann constant).

It is useful to define the phase-space vector $\overrightarrow{r}(t) = \left( \overrightarrow{q}, \mathbb{M}^{-1}\overrightarrow{p} \right)^\mathsf{T}$ (note that $\overrightarrow{v} = \mathbb{M}^{-1}\overrightarrow{p}$ is just the velocity vector) so the equations of motion for this vector are
\begin{equation}
  \dot{\overrightarrow{r}}(t) = \mathbb{A} \, \overrightarrow{r}(t) + \mathbb{L}\overrightarrow{\xi}(t),
  \label{eq:vectorEqOfMotion}
\end{equation}
with
\begin{align}
  \mathbb{A} &=
  \left(
  \begin{array}{cc}
    \mathbb{0}_{2 \times 2} & \mathbb{1}_{2 \times 2}
    \\
    -\mathbb{M}^{-1}\mathbb{K} & -\mathbb{M}^{-1}\Gamma
  \end{array}
  \right),
  \nonumber
  \\
  \mathbb{L} &=
  \left(
  \begin{array}{c}
    \mathbb{0}_{2\times 2} \\ \mathbb{M}^{-1}
  \end{array}
  \right),
\end{align}
and $\overrightarrow{\xi}(t) = \left( \xi_L(t),\xi_R(t) \right)^\mathsf{T}$, $\Gamma = diag(\gamma_L,\gamma_R)$. $\mathbb{0}_{n\times n}$ and $\mathbb{1}_{n\times n}$ are the $n$-th dimensional squared 0 matrix and identity matrix respectively. With the vector notation the correlation of the white-noise forces can be written as
\begin{equation}
  \expval{\overrightarrow{\xi}(t)\overrightarrow{\xi}(t')^\mathsf{T}} = 2 \mathbb{D}\delta(t-t'),
\end{equation}
with $\mathbb{D} = diag(D_L,D_R)$.
\section{Covariance matrix in the steady state\label{sec:covMatrix}}
We define the covariance matrix of the system as $\mathbb{C}(t) = \expval{\overrightarrow{r}(t)\overrightarrow{r}(t)^\mathsf{T}}$. This matrix is important because the heat transport properties can be extracted from it. In particular, the kinetic temperatures of the masses, $T_1(t)$ and  $T_2(t)$, are
\begin{align}
  T_1(t) &= \frac{\expval{ p_1^2(t)}}{m_1 k_B} = \frac{m_1 C_{3,3}(t)}{k_B},
  \nonumber\\
   T_2(t) &= \frac{\expval{ p_2^2(t)}}{m_2 k_B} = \frac{m_2 C_{4,4}(t)}{k_B}.
  \label{eq:Temperature_definition}
\end{align}
One approach to find the covariance matrix is to solve Eq. \eqref{eq:vectorEqOfMotion}. However, this requires solving the equations explicitly or simulate them numerically many times to find the covariance matrix for the ensemble of simulated stochastic trajectories. Instead, we proceed by looking for an ordinary differential equation that gives the evolution of the covariance matrix as described in \cite{Sarkka2019,Rieder1967,Casher1971}. Differentiating $\mathbb{C}(t)$ with respect to time and using Eq. \eqref{eq:vectorEqOfMotion} we get
\begin{align}
  \frac{d}{dt}\mathbb{C}(t) &=
  \mathbb{A}\mathbb{C}(t) +
  \mathbb{C}(t) \mathbb{A}^\mathsf{T}
  \nonumber\\
  &+
  \mathbb{L}\expval{ \overrightarrow{\xi}(t)\overrightarrow{r}(t)^\mathsf{T}}
  \nonumber\\
  &+
  \expval{ \overrightarrow{r}(t)\overrightarrow{\xi}(t)^\mathsf{T}}\mathbb{L}^\mathsf{T}.
  \label{eq:evolutionOfCovariances}
\end{align}
The solution of Eq. \eqref{eq:evolutionOfCovariances} allows us to find the local temperatures of the masses as a function of the bath temperatures (Eq. \eqref{eq:Temperature_definition}) at all times. In particular, we are interested in the covariance matrix in the steady state, i.e., for $t\to \infty$. According to the Novikov Theorem \cite{Novikov1965} we can write down the covariance matrix in the steady state without having to integrate the differential equation. We now show how to get the steady-state covariance matrix.

In the steady state, the covariance matrix is constant ($\frac{d}{dt}\mathbb{C}(t)=0$), therefore it satisfies
\begin{align}
  &\mathbb{A}\mathbb{C}^{s.s.} +
  \mathbb{C}^{s.s.} \mathbb{A}^\mathsf{T}=
  \nonumber\\
  &- \mathbb{L}\expval{ \overrightarrow{\xi}\overrightarrow{r}^\mathsf{T}}^{s.s.}
  - \expval{ \overrightarrow{r}\overrightarrow{\xi}^\mathsf{T}}^{s.s.}\mathbb{L}^\mathsf{T},
  \label{eq:SteadyStateEquation_raw}
\end{align}
with $\small\{\cdot\small\}^{s.s.}\equiv \lim\limits_{t \to \infty} \small\{\cdot\small\}(t)$. Equation \eqref{eq:SteadyStateEquation_raw} is an algebraic equation whose solution is the steady-state covariance matrix $\mathbb{C}^{s.s.}$. However, the two terms $\expval{ \overrightarrow{\xi}\overrightarrow{r}^\mathsf{T}}^{s.s.}$ and  $\expval{\overrightarrow{r}\overrightarrow{\xi}^\mathsf{T}}^{s.s.}$ need to be calculated before working out the solution. One approach to calculate $\expval{\overrightarrow{\xi}\overrightarrow{r}^\mathsf{T}}^{s.s.}$ would be to solve Eq. \eqref{eq:vectorEqOfMotion}, but this is exactly what we are trying to avoid. It is here when the Novikov theorem comes useful, since it lets us compute $\expval{ \overrightarrow{\xi}\overrightarrow{r}^\mathsf{T}}^{s.s.}$ without having to integrate the equations of motion. Using this theorem and the $\delta$-correlation of the noises, we find the $ij$-th component of $\expval{ \overrightarrow{\xi}(t)\overrightarrow{r}(t)^\mathsf{T}}$,
\begin{align}
  \expval{ \xi_i(t) r_j(t) } &= \sum_{k=1}^2 \int_0^t d\tau\,\expval{ \xi_i(t) \xi_k(\tau)}
  \,
  \expval{ \frac{\delta r_j(t)}{\delta \xi_k(\tau)} }\nonumber
  \\
  &= \sum_{k=1}^2 \mathbb{D}_{ik}
  \,
  \lim_{\tau \to t^{-}}
  \,
  \expval{ \frac{\delta r_j(t)}{\delta \xi_k(\tau)} },
\end{align}
where $\lim\limits_{\tau \to t^{-}}$ is the limit when $\tau$ goes to $t$ from below. Evaluation of the functional derivative ${\delta r_j(t)}/{\delta \xi_k(\tau)}$ for the $\tau \to t^{-}$ limit gives
\begin{equation}
  \expval{ \overrightarrow{\xi}(t)\overrightarrow{r}(t)^\mathsf{T}} = \mathbb{D}\mathbb{L}^\mathsf{T}.
\end{equation}
Now, the algebraic equation that gives the steady-state covariance matrix becomes
\begin{equation}
  \mathbb{A}\mathbb{C}^{s.s.} +
  \mathbb{C}^{s.s.}\mathbb{A}^\mathsf{T}
  =
  -\mathbb{B},
  \label{eq:SteadyStateEquation}
\end{equation}
with $\mathbb{B} = 2 \mathbb{L}\mathbb{D}\mathbb{L}^\mathsf{T}$. By definition, the covariance matrix is  symmetric, but there are also  additional restrictions imposed by the equations of motion and the steady-state condition, which reduce the dimensionality of the problem of solving Eq. \eqref{eq:SteadyStateEquation} \cite{Simon2019}. Since ${d \expval{ q_i q_j }}/{dt} = 0$ in the steady state, we have
\begin{align}
  \expval{ p_1 q_1}^{s.s.} &= \expval{ p_2 q_2}^{s.s.} = 0,\nonumber\\
  \frac{\expval{ p_1 q_2}^{s.s.}}{m_1}&=-\frac{\expval{ q_1 p_2}^{s.s.}}{m_2}.
  \label{eq:ExtraConditionSteadyState}
\end{align}
Taking \eqref{eq:ExtraConditionSteadyState} into account, the steady-state covariance matrix takes the form
\begin{equation}
  \begin{split}
    \mathbb{C}^{s.s.} =
    \left(
    \begin{array}{cccc}
      \expval{ q_1^2}^{s.s.}  & \expval{ q_1 q_2}^{s.s.}  & 0 & \frac{\expval{ p_2 q_1}^{s.s.} }{m_2} \\
      \expval{ q_1 q_2}^{s.s.}  & \expval{ q_2^2}^{s.s.}  & -\frac{\expval{ p_2 q_1}^{s.s.} }{m_2} & 0 \\
      0 & -\frac{\expval{ p_2 q_1}^{s.s.} }{m_2} & \frac{\expval{ p_1^2}^{s.s.} }{m_1^2} & \frac{\expval{ p_1 p_2}^{s.s.} }{m_1 m_2} \\
      \frac{\expval{ p_2 q_1}^{s.s.} }{m_2} & 0 & \frac{\expval{ p_1 p_2}^{s.s.} }{m_1 m_2} & \frac{\expval{ p_2^2}^{s.s.} }{m_2^2} \\
      \end{array}
      \right)
    \end{split}
    \label{eq:steadyStateCovarianceMatrix}\,.
\end{equation}
The explicit set of equations for the components of $\mathbb{C}^{s.s}$ can be found in Appendix \ref{AppStationaryStateEquations}.
\section{Solutions\label{sec:solutions}}
In this section we use the solution to Eq. \eqref{eq:SteadyStateEquation} to write down the temperatures and currents in the steady state. We use Mathematica to obtain analytic expressions for the temperatures,
\begin{align}
  T_1 &= \frac{T_L \mathcal{P}_{1,L}(k) + T_R \mathcal{P}_{1,R}(k)}{\mathcal{D}(k)},\nonumber
  \\
  T_2 &= \frac{T_L \mathcal{P}_{2,L}(k) + T_R \mathcal{P}_{2,R}(k)}{\mathcal{D}(k)},
  \label{eq:ModelBTemperatures}
\end{align}
where $\mathcal{D}(k) =  \sum\limits_{n=0}^2 \mathcal{D}_n k^n$ and $\mathcal{P}_{i,(L/R)}(k) = \sum\limits_{n=0}^2 a_{i,n,(L/R)} k^n$ are polynomials in the coupling constant $k$ with coefficients
\begin{widetext}
  \begin{align}
    \mathcal{D}_0 &= a_{1,0,L} = a_{2,0,R} = \gamma _L \gamma _R \left[h^{(1)} \left(\gamma_L k_R +\gamma_R k_L \right)+\left(m_1 k_R-m_2 k_L\right)^2\right],\nonumber
    \\
    \mathcal{D}_1 &= a_{1,1,L} = a_{2,1,R} = \gamma _L \gamma _R \left[h^{(0)} h^{(1)}+2 \left(m_1-m_2\right) \left(m_1 k_R-m_2 k_L\right)\right],\nonumber
    \\
    \mathcal{D}_2 &= h^{(0)} h^{(2)},\nonumber
    \\
    a_{1,2,L} &= \gamma _L \left(m_2 h^{(1)} + \gamma_R (m_1 - m_2)^2 \right),\nonumber
    \\
    a_{1,2,R} &= h^{(1)} m_1 \gamma_R,\nonumber
    \\
    a_{2,2,L} &= h^{(1)} m_2 \gamma_L,\nonumber
    \\
    a_{2,2,R} &= \gamma _R \left( m_1 h^{(1)} + \gamma_L (m_1-m_2)^2 \right),\nonumber
    \\
    a_{1,0,R} &= a_{1,1,R} = a_{2,0,L} = a_{2,1,L} = 0,
    \label{eq:SolutionPolynomialCoefficients}
  \end{align}
\end{widetext}
where $h^{(n)}\equiv \gamma_R m_1^n + \gamma_L m_2^n$. The currents from the baths to the masses \cite{Simon2019} are
\begin{equation}
  \begin{split}
    J_L &= k_B \frac{\gamma_L}{m_1} \left( T_L - T_1 \right),\\
    J_R &= k_B \frac{\gamma_R}{m_2} \left( T_R - T_2 \right),
    \label{eq:currents_definition}
  \end{split}
\end{equation}
\\
with $T_i$ given by Eq. \eqref{eq:ModelBTemperatures}. Since, in the steady state, $J_L = -J_R$ we will use the shorthand notation $J \equiv J_L$. Substituting Eq. \eqref{eq:ModelBTemperatures} into Eq.  \eqref{eq:currents_definition} we get for the heat current
%
%
\begin{equation}
  J = \kappa\;(T_L - T_R),
  \label{eq:CurrentsInModelB}
\end{equation}
where $\kappa = k_B {k^2\gamma_L \gamma_R h^{(1)}}/{\mathcal{D}(k)}$ acts as an effective thermal conductance, which depends on the parameters of the system, i.e., the masses and spring constants, and also on the friction coefficients of the baths. From Eq. \eqref{eq:CurrentsInModelB} it could be thought that inverting the temperatures of the baths would only lead to an exchange of heat currents. However, since the thermal conductance $\kappa$ depends on the friction coefficients, the exchange of the baths implies a change in its value. Moreover, it is possible to have temperature-dependent friction coefficients, as it happens in the physical set-up of laser-cooled trapped ions described in Section \ref{sec:TrappedIonSetUp}.

\section{Relation of the Model to a trapped ion set-up \label{sec:TrappedIonSetUp}}

As we mentioned, the parameters $k$, $k_L$ and $k_R$ can be related to the elements of the Hessian matrix of a system in a stable equilibrium position. In this section we will identify these parameters with the Hessian matrix of a pair of trapped ions. Here we consider two different set-ups: two ions in a collective trap, and two ions in individual traps. In Section \ref{sec:lookingForR} we focus on two ions in individual traps to illustrate the analysis of rectification.

In both set-ups we assume strong confinement in the radial direction, making the effective dynamics one-dimensional. We will also assume that the confinement in the axial direction is purely electrostatic, which makes the effective spring constant independent of the mass of the ions \cite{Leibfried2003}. Additionally, we will relate the temperatures and friction coefficients of the Langevin baths to those corresponding to Doppler cooling.

\subsection{Collective trap}

Consider two ions of unit charge with masses $m_1$ and $m_2$ trapped in a collective trap. Assuming strong radial confinement and purely electrostatic axial confinement, both ions feel the same harmonic oscillator potential with trapping constant $k_{trap}$ \cite{Leibfried2003}. The potential describing the system is
\begin{equation}
  V_{collective} = \frac{1}{2}k_{trap} \left( x_1^2 + x_2^2\right) + \frac{\mathcal{C}}{x_2-x_1},
\end{equation}
with $\mathcal{C}=\frac{Q^2}{4\pi\varepsilon_0}$. The equilibrium positions for this potential are
\begin{equation}
  x_2^{eq} = -x_1^{eq} =
  \label{eq:equilibriumPositionsCollectiveTrap}\left(\frac{1}{2}\right)^{2/3} \left(\frac{Q^2}{4\pi\varepsilon_0 k_{trap}}\right)^{1/3}.
\end{equation}
Assuming small oscillations of the ions around the equilibrium positions, the Hessian matrix of the system is
\begin{align}
  \mathbb{K}_{1,2} &= -\frac{Q^2}{2\pi\varepsilon_0}\frac{1}{(x_2^{eq}-x_1^{eq})^3} = -k_{trap},\nonumber
  \\
  \mathbb{K}_{1,1} &= k_{trap} + \frac{Q^2}{2\pi\varepsilon_0}\frac{1}{(x_2^{eq}-x_1^{eq})^3} = 2 k_{trap},\nonumber
  \\
  \mathbb{K}_{2,2} &= k_{trap} + \frac{Q^2}{2\pi\varepsilon_0}\frac{1}{(x_2^{eq}-x_1^{eq})^3} = 2 k_{trap}.
  \label{eq:HessianOffDiagonalCollective}
\end{align}
Using Eq. \eqref{eq:HessianOffDiagonalCollective} we can relate the parameters of this physical set-up to those of the model described in Section \ref{sec:Physical_Model} to find
\begin{equation}
  k_L = k_R = k = k_{trap}.
\end{equation}

\subsection{Individual on-site traps}

We can make the same assumptions for the axial confinement as in the previous subsection but now each of the ions is in an individual trap with spring constants $k_{trap,L}$ and $k_{trap,R}$ respectively. The potential of the system is
\begin{align}
    V_{individual} &= \frac{1}{2}k_{trap,L}\left(x_1 -x_L\right)^2 +\frac{1}{2}k_{trap, R}\left(x_2 -x_R\right)^2 \nonumber \\&+ \frac{\mathcal{C}}{x_2-x_1},
\end{align}
where $x_L$ and $x_R$ are the center positions of the on-site traps. The elements of the Hessian matrix in the equilibrium position are
\begin{align}
  \mathbb{K}_{1,2} &= -\frac{Q^2}{2\pi\varepsilon_0}\frac{1}{(x_2^{eq}-x_1^{eq})^3},\nonumber
  \\
  \mathbb{K}_{1,1} &= k_{trap,L} + \frac{Q^2}{2\pi\varepsilon_0}\frac{1}{(x_2^{eq}-x_1^{eq})^3},\nonumber
  \\
  \mathbb{K}_{2,2} &= k_{trap,R} + \frac{Q^2}{2\pi\varepsilon_0}\frac{1}{(x_2^{eq}-x_1^{eq})^3}.
  \label{eq:HessianOffDiagonalOnSite}
\end{align}
Comparing the parameters in Eq. \eqref{eq:HessianOffDiagonalOnSite} with those in the model described in Section \ref{sec:Physical_Model} we identify
\begin{align}
  k_L &= k_{trap,L},\nonumber\\
  k_R &= k_{trap,R},\nonumber\\
  k &= \frac{Q^2}{2\pi\varepsilon_0}\frac{1}{(x_2^{eq}-x_1^{eq})^3}\,.
\end{align}
In this case, the analytic expressions for the equilibrium positions are more complicated. We get for the distance between the equilibrium positions of the ions
\begin{align}
  &(x_2 - x_1)^{(eq)} = \frac{1}{3} \Delta x_{LR}\nonumber\\
  &- \frac{1}{6}\Big[ \frac{2^{2/3}\zeta}{k_{trap,L} k_{trap,R} (k_{trap,L} + k_{trap,R})}\nonumber\\
  &+ \frac{2^{4/3} k_{trap,L} k_{trap,R} (k_{trap,L} + k_{trap,R}) (x_R-x_L)^2}{\zeta} \Big]\,,
\end{align}
where $\Delta x_{LR} = (x_R-x_L)$ and $\zeta = \left( Y - \eta \right)^{(1/3)}$, with
\begin{align}
  &Y = 3 \sqrt{3} \bigg\{\mathcal{C} k_{trap,L}^4 k_{trap,R}^4 \left(k_{trap,L}+k_{trap,R}\right)^{7}\times\nonumber\\& \quad\quad \left[4 k_{trap,L} k_{trap,R} \Delta x_{LR}^3+27 \mathcal{C} \left(k_{trap,L}+k_{trap,R}\right)\right]\bigg\}^{(1/2)},\nonumber
  \\
  &\eta =  k_{trap,L}^2 k_{trap,R}^2 \left(k_{trap,L}+k_{trap,R}\right)^{3}\times\nonumber\\ &\quad\quad\left[2 k_{trap,L} k_{trap,R} \Delta x_{LR}^3+27 \mathcal{C} \left(k_{trap,L}+k_{trap,R}\right)\right]\,.
\end{align}
In this set-up, the coupling between the ions $k$ can be controlled by changing the distance between the on-site traps.

\subsection{Optical molasses and Langevin baths}

Trapped ions may be cooled down by a pair of counterpropagating lasers which are red-detuned with respect to an internal atomic transition of the ions. This technique is known as Doppler cooling or optical molasses \cite{Chu1985,Cohen1992,Metcalf1999,Metcalf2003}. The off-resonant absorption of laser photons by the ions exerts a damping-like force that slows them down. The spontaneous emission of the ions produces heating due to the random recoil generated by the emitted photons. Both, the friction and recoil force are in balance, and eventually the ion thermalizes to a finite temperature.
Thus the effect of the lasers on the ion is equivalent to a Langevin bath with temperature $T_{molass}$ and friction coefficient $\gamma_{molass}$. The temperature and friction coefficients are controlled with the laser intensity $I$ and frequency detuning $\delta$ with respect to the selected internal transition by the expressions \cite{Cohen1992,Metcalf2003,Ruiz2014},
\begin{align}
  \gamma_{molass}(I,\delta) &= -4 \hbar \left(\frac{\delta + \omega_0}{c}\right)^2 \left(\frac{I}{I_0}\right)\frac{2\delta/\Gamma}{\left[1 + (2\delta/\Gamma)^2\right]^2},\nonumber\\
  T_{molass}(\delta) &= -\frac{\hbar \Gamma}{4 k_B} \frac{1+(2\delta/\Gamma)^2}{(2\delta/\Gamma)},
  \label{eq:DopplerCooling}
\end{align}
where $\omega_0$ is the frequency of the selected internal atomic transition, $\Gamma$ is the natural width of the excited state, and $I_0$ is the saturation intensity.
\section{Looking for rectification\label{sec:lookingForR}}
%
%
%
We will say that we observe rectification whenever the heat current $J$ for a configuration of the baths changes when we exchange the baths to $\tilde{J}$. The important point here is to define what is  meant by \textit{exchanging the baths}. We consider that a bath is characterized, not only by its temperature $T$ but also by its coupling  to the system by means of the friction coefficient $\gamma$, so, exchanging the baths is achieved by exchanging both the temperatures and the friction coefficients, as summarized in Table \ref{tab:reversed_bath}.

When implementing temperatures and friction coefficients by lasers, this exchange operation is performed by changing the values of the intensities and detunings acting on each ion (Eq. \eqref{eq:DopplerCooling}). The exchange operation is straightforward when the two ions are either of the same species or isotopes of each other, since the only required action is to exchange the values of the detunings of the lasers without modifying the intensities. However, if we deal with two different species, i.e., with two different atomic transitions, the laser wavelengths and the decay rates  depend on the species. Then, exchanging the temperatures by modifying the detunings, keeping the laser intensities constant, does not necessarily imply an exchange of the friction coefficients. Nevertheless it is possible to adjust the laser intensities so that the friction coefficients get exchanged and that is the assumption hereafter. The idea of implementing a bath exchange like this follows the same line of thought as \cite{Pereira2017}, since we are adding a temperature dependent feature to the system -the friction coefficients- that changes as the baths are inverted.

\begin{figure}
  \includegraphics[width=\linewidth]{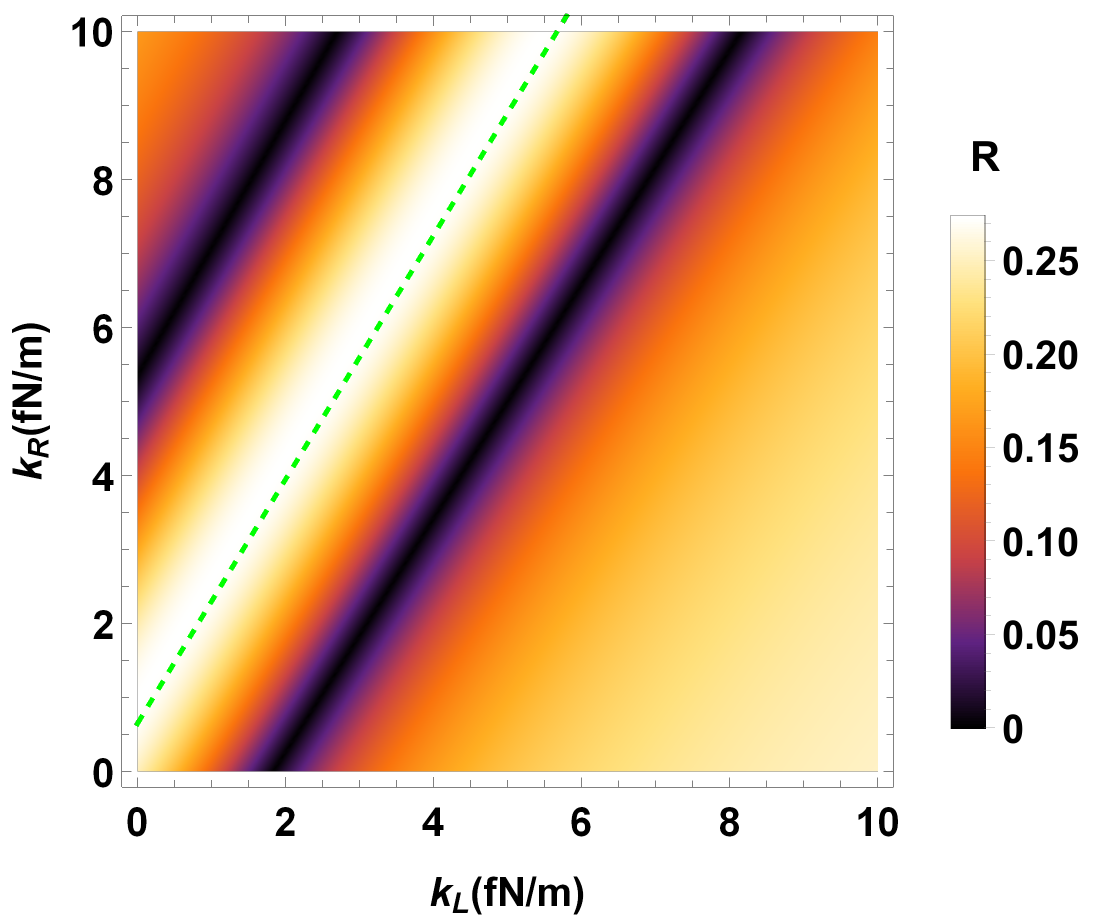}
  \caption{Rectification, $R$, in the $k_L k_R$ plane for $k = 1.17 \times$ fN/m, $\gamma_L = 6.75\times 10^{-22}$ kg/s, and $\gamma_R = 4.64\gamma_L$.}
  \label{fig:Fig_rectification_K_plane}
\end{figure}

To measure rectification, we will use the rectification coefficient $R$ defined as
\begin{equation}
  R = \frac{\abs{J-\tilde{J}}}{\max(J,\tilde{J})},
  \label{eq:Rectification}
\end{equation}
that is, the ratio between the difference of heat currents and the largest one. As defined, $R=0$ for no asymmetry of the heat currents and $R=1$ when they are maximally asymmetric.

\begin{table}[]
\caption{Definition of forward and reversed (exchanged) bath configurations.}
\begin{tabular}{lcc}
\hline
                 & forward                & reversed                                                       \\ \hline
Bath Friction    & $\gamma_L$, $\gamma_R$ & $\tilde{\gamma}_L =\gamma_R $,  $\tilde{\gamma}_R =\gamma_L $   \\
Bath Temperature & $T_L$, $T_R$           & $\tilde{T}_L =T_R $,  $\tilde{T}_R =T_L $                     \\
\hline
\end{tabular}
\label{tab:reversed_bath}
\end{table}

\subsection{Parametric exploration}

We have explored thoroughly the space formed by the parameters of the model to find asymmetric heat transport, namely, $m_1,m_2,k,k_L,k_R,\gamma_L,\gamma_R$. We have fixed the values of some of the parameters to realistic ones while we have varied the rest. We have set the masses to $m_1 = 24.305$ a.u. and $m_2 = 40.078$ a.u., which correspond to Mg and Ca, whose ions are broadly used in trapped-ion physics. The temperatures are also fixed and, as Eq. \eqref{eq:CurrentsInModelB} shows, rectification does not formally depend on the temperature in this model, unless we set the friction coefficients as a function of temperature using Eq. \eqref{eq:DopplerCooling} explicitly.

Figure \ref{fig:Fig_rectification_K_plane} depicts the values of the rectification after sweeping the $k_L k_R$ plane for fixed values of $k$, $\gamma_L$, and $\gamma_R$. A remarkable result from this figure is that parallel lines appear alternating minima and maxima of $R$. With a numerical fitting, we find that the line corresponding to the highest maximum value of $R$ is determined by
\begin{equation}
  \frac{k+k_L}{m_1} = \frac{k+k_R}{m_2}.
  \label{eq:MaxRLines}
\end{equation}
In a trapped-ion context the condition \eqref{eq:MaxRLines} may be imposed by adjusting the distance of the traps for fixed $k_L$ and $k_R$. It is also  remarkable that when Eq. \eqref{eq:MaxRLines} is satisfied, the rectification no longer depends on the spring constants of the model. This last result can be  found assuming  Eq. \eqref{eq:MaxRLines} when calculating the currents with Eq. \eqref{eq:CurrentsInModelB} and $R$ with Eq. \eqref{eq:Rectification},
\begin{equation}
    R=
    \begin{cases}
      1-\frac{a+g}{1+ag} &\text{ if }(a+g)<(1+ag)\\
      1-\frac{1+ag}{a+g} &\text{ if }(a+g)>(1+ag)\\
      0 &\text{ if } (a+g)=(1+ag)\,,
    \end{cases}
  \label{eq:maxRExpression}
\end{equation}
where $a$ and $g$ are the mass and friction coefficients ratios
\begin{align}
  a &= m_2/m_1,\nonumber\\
  g &= \gamma_R/\gamma_L.
\end{align}
The maximal rectification found does not scale with the magnitude of the masses or the friction coefficients, just with their ratios. Besides a high $R$, it is important to have non-vanishing heat currents
\cite{Simon2019}. Using again  Eq. \eqref{eq:MaxRLines} in the expression for the currents \eqref{eq:CurrentsInModelB}, the maximum current $J_{\max} = \max(\big|{J}\big|,\big|\tilde{J}\big|)$ is
\begin{align}
    &J_{\max}=\begin{cases}
   \frac{k_B g\gamma_L k^2 \abs{T_L-T_R}}{(a+g)(g\gamma_L^2(k_L+k)+k^2m_1)} & \text{ if }(a+g)<(1+ag)
    \\
    \frac{k_B g\gamma_L k^2 \abs{T_L-T_R}}{(1+ag)(g\gamma_L^2(k_L+k)+k^2m_1)}& \text{ if }(a+g)>(1+ag)\,.
    \end{cases}
    \label{eq:maxJExpression}
\end{align}
Now we analyze how the parameters $a$ and $g$ affect the maximum current $J_{max}$ in \eqref{eq:maxJExpression}. To do this, we can divide the $ag$ plane in four quadrants by the axes $a = 1$ and $g = 1$ (in those axes $R = 0$). In Eq. \eqref{eq:maxJExpression} the parameter $a$ appears only in the denominator, thus for a higher $a$, a smaller current is found. The quadrants with $a < 1$ will be better for achieving large currents. However, $g$ appears both in the numerator and denominator so there is no obvious advantageous quadrant for this parameter.

Equation \eqref{eq:maxRExpression} is symmetric upon the transformations $a \leftrightarrow 1/a$ and $g \leftrightarrow 1/g$. Using a logarithmic scale for $a$ and $g$, the resulting $R$ map will be symmetric with respect to the $a=1$ and $g=1$ axes. We can limit ourselves to analyze the quadrant $a > 1$, $g > 1$, as the results in other quadrants will be equivalent upon transformations $a \leftrightarrow 1/a$ and $g \leftrightarrow 1/g$.

\begin{figure}
  \includegraphics[width=\linewidth]{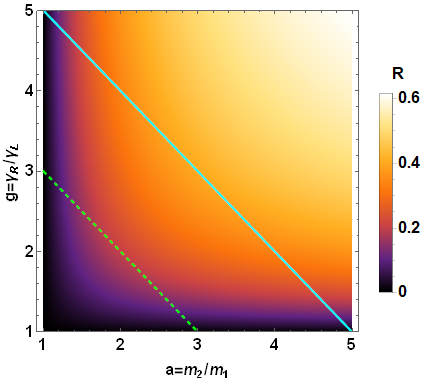}
  \caption{Rectification factor, $R$, given by Eq. \eqref{eq:maxRExpression}.}
  \label{fig:R_g_a_plane}
\end{figure}

Fig. \ref{fig:R_g_a_plane} shows the rectification given by Eq. \eqref{eq:maxRExpression} in terms of $a$ and $g$. Along any diagonal line (parallel to the solid cyan or the dashed green lines), the maximum value is at the center, that is, when $a = g$. However, if we fix $a$, increasing $g$ always increases $R$. Although we could increase $g$ arbitrarily to get more rectification this is not a realistic option in a trapped-ion set-up. Since $g$ is defined as the ratio between the friction coefficients, increasing it means making either $\gamma_L$ go to 0 or $\gamma_R$ to infinity. Making $\gamma_L$ go to 0 decouples one of the ions from the bath, so the heat current tends to vanish in any direction. Also, increasing $\gamma_R$ arbitrarily is impossible since the Doppler cooling friction coefficient as a function of the laser detuning (Eq. \eqref{eq:DopplerCooling}) is bounded. Although Eq. \eqref{eq:DopplerCooling} suggests that boosting the laser intensity can also increase the friction coefficient, this is not an option since Eq. \eqref{eq:DopplerCooling} is just an approximation for low laser intensities. When going to higher intensities, the emission/absorption of photons by the ion is saturated and the friction coefficient reaches a finite value proportional to the width $\Gamma$ of the excited state \cite{Metcalf2003}. As a compromise between feasibility and high $R$, we set the ratio between the friction coefficients $g$ to be equal to the mass ratio $a$. As shown  in Fig. \ref{fig:R_g_a_plane}, along the solid-cyan and dashed-green diagonal lines the maximum $R$ is achieved for $a = g$. Fig. \ref{fig:Fig_PerfectRectification} shows the rectification in Eq. \eqref{eq:maxRExpression} for the line $a = g$. When both parameters are large enough, the rectification goes to 1.
\subsection{Spectral match/mismatch approach to rectification}
\begin{figure}
  \includegraphics[width=\linewidth]{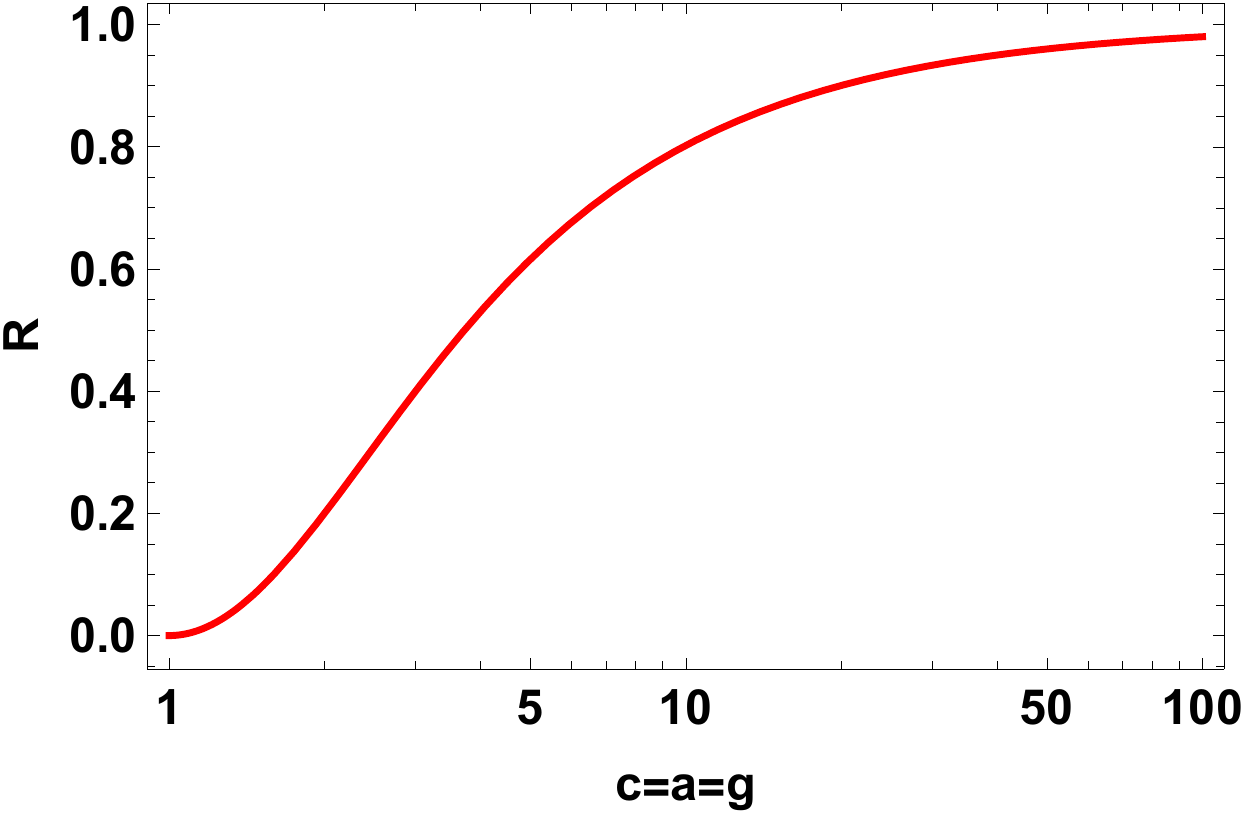}
  \caption{Rectification for different values of $c=m_2/m_1=\gamma_R/\gamma_L$ when the maximum condition in the $k_L k_R$ plane is satisfied (Eq. \eqref{eq:MaxRLines}).}
  \label{fig:Fig_PerfectRectification}
\end{figure}

The match/mismatch between the power spectra of the particles controls the heat currents in the system \cite{Terraneo2002,Li2004}. A good match between the power spectra of the two ions in a large range of frequencies yields a higher heat current through the system while the mismatch  reduces the heat current.
If there is a good match between the spectra of the ions (i.e., their peaks overlap in a broad range of frequencies) for a certain baths configuration, and mismatch when the baths exchange, the system will present heat rectification.

We have studied the phonon spectra of our model for several sets of parameters exhibiting no rectification or strong rectification. The phonon spectra of the ions is calculated through the spectral density matrix. For a real-valued stochastic process $\overrightarrow{x}(t)$, its spectral density matrix is defined as \cite{Sarkka2019}
\begin{equation}
  \mathbb{S}_{\overrightarrow{x}}(\omega) \equiv \expval{ \overrightarrow{X}(\omega) \overrightarrow{X}^\mathsf{T}(-\omega) },
  \label{eq:SpectralDensityDefinition}
\end{equation}
with $\overrightarrow{X}(\omega)$ being the Fourier transform of $\overrightarrow{x}(t)$ (we are using the convention of multiplying by a factor of $1$ and $\frac{1}{2\pi}$ for the transform and its inverse operation). A justification of the use of the spectral density matrix to understand heat transport arises from the Wiener-Khinchin theorem \cite{Sarkka2019}, which says that the correlation matrix of a stationary stochastic process in the steady state is the inverse Fourier transform of its spectral density matrix $\expval{\overrightarrow{r}(t)\overrightarrow{r}^\mathsf{T}(t+\tau)} = \mathcal{F}^{-1}[\mathbb{S}_{\overrightarrow{r}}(\omega)](\tau)$. This result allows us to write down the covariance matrix in the steady state through the spectral density as
\begin{equation}
  \mathbb{C}^{s.s.} = \frac{1}{2\pi} \int_{-\infty}^{\infty}d\omega\;\mathbb{S}_{\overrightarrow{r}}(\omega).
  \label{eq:Wiener-Khinchin}
\end{equation}
Eq. \eqref{eq:Wiener-Khinchin} directly connects the spectral density matrix to the steady-state temperature and, therefore, to the heat currents (in Section \ref{sec:covMatrix} we saw that  $T_1^{s.s.} = {m_1 C_{3,3}^{s.s.}}/{k_B}$ and $T_2^{s.s.} = {m_2 C_{4,4}^{s.s.}}/{k_B}$).

\begin{figure}[t]
  \includegraphics[width=\linewidth]{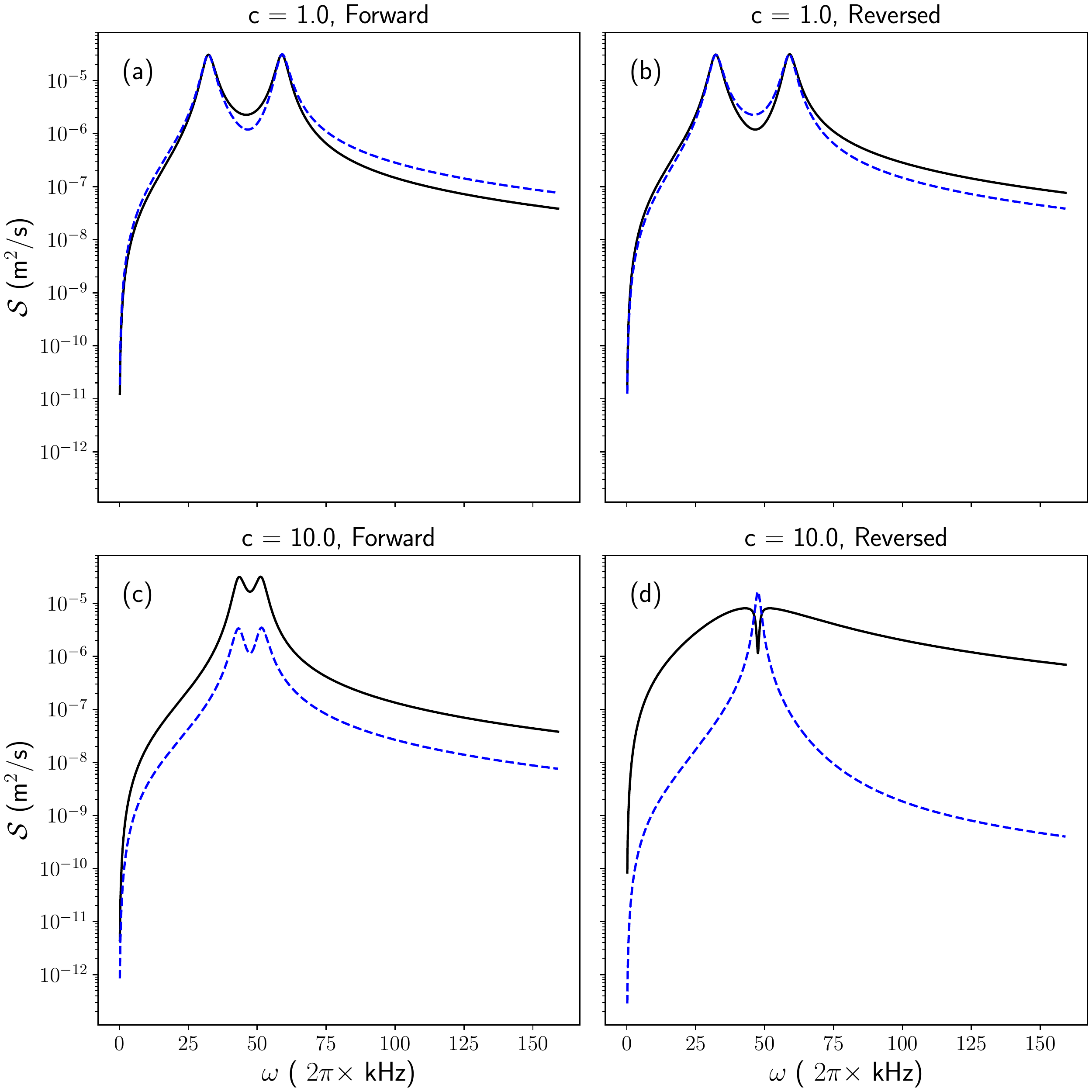}
  \caption{Spectral densities of the velocities of the ions ($r_3$ and $r_4$) corresponding to different values of $c$ in Fig. \ref{fig:Fig_PerfectRectification}: (a), (b) for $c=1$ and (c), (d) for $c=10$. Solid, black lines correspond to the left ion velocity spectral density $\mathbb{S}_{3,3}(\omega)$ and dashed, blue lines correspond to the right ion velocity spectral density $\mathbb{S}_{4,4}(\omega)$. (a) and (b) correspond to $R = 0$:  the overlap between the phonon bands is the same in the forward and reversed configurations. (c) and (d) correspond to $R\approx 0.8$:  in the forward configuration (c)  the phonons match better than in the reversed configuration (d).}
  \label{fig:Figure_Spectra}
\end{figure}

For the vector process $\overrightarrow{r}(t)$ describing the evolution of our system we have $\overrightarrow{R}(\omega) = \left( i \omega - \mathbb{A} \right)^{-1}\mathbb{L}\overrightarrow{\Xi}(\omega)$ with $\overrightarrow{\Xi}(\omega)$ being the Fourier transform of the white noise $\overrightarrow{\xi}(t)$. Note that $\overrightarrow{\Xi}(\omega)$ does not strictly exist, because it is not square-integrable, however its spectral density is $\mathbb{S}_{\overrightarrow{\xi}}(\omega) = 2 \mathbb{D}$ \cite{Sarkka2019}, which is flat as expected for a white noise. Therefore, the spectral density matrix of the system is
\begin{equation}
  \mathbb{S}_{\overrightarrow{r}} = 2 \left(  \mathbb{A} - i\omega\right)^{-1}\mathbb{L}\mathbb{D}\mathbb{L}^\mathsf{T}\left(  \mathbb{A} + i\omega\right)^{-\mathsf{T}}.
  \label{eq:SpectralDensityToyModelB}
\end{equation}
As we can see in Eq. \eqref{eq:SpectralDensityToyModelB}, the imaginary part of the eigenvalues of the dynamical matrix $\mathbb{A}$ correspond to the peaks in the spectrum whereas the real part dictates their width. The spectral density matrix of our model is
\begin{equation}
  \mathbb{S}_{\overrightarrow{r}}(\omega) = 2 k_B \frac{\gamma_L T_L\mathbb{S}_L(i\omega)+\gamma_L T_R\mathbb{S}_R(i\omega)}{(m_1 m_2)^2 P_\mathbb{A}(i\omega)P_\mathbb{A}(-i\omega)},
\end{equation}
where $P_\mathbb{A}(\lambda)$ is the characteristic polynomial of the dynamical matrix $\mathbb{A}$ and $\mathbb{S}_L(\omega)$, $\mathbb{S}_R(\omega)$ are the matrix polynomials in the angular frequency $\omega$ whose coefficients are defined in Appendix \ref{AppSpecDenMat}. Equation \eqref{eq:SpectralDensitiesVelocities} gives the full expressions of the spectral densities for the velocities, $\mathbb{S}_{3,3}(\omega) = \expval{R_3(\omega)R_3(-\omega)}$ for the left ion, and $\mathbb{S}_{4,4}(\omega) = \expval{R_4(\omega)R_4(-\omega)}$ for the right ion, since they are the elements related to the calculation of the heat current using Eq. \eqref{eq:Wiener-Khinchin},
\begin{widetext}
  \begin{align}
    \mathbb{S}_{3,3}(\omega) &= 2 k_B \frac{\gamma_R k^2 T_R \omega ^2+\gamma_L T_L \left[\omega ^4 \left(\gamma_R^2-2 k m_2-2 k_R m_2\right)+\omega ^2 (k+k_R)^2+m_2^2 \omega ^6\right]}{(m_1 m_2)^2 P_\mathbb{A}(i\omega)P_\mathbb{A}(-i\omega)},\nonumber\\
    %
    %
    \mathbb{S}_{4,4}(\omega) &= 2 k_B \frac{\gamma_L k^2 T_L \omega ^2+\gamma_R T_R \left[\omega ^4 \left(\gamma_L^2-2 k m_1-2 k_L m_1\right)+\omega ^2 (k+k_L)^2+m_1^2 \omega ^6\right]}{(m_1 m_2)^2 P_\mathbb{A}(i\omega)P_\mathbb{A}(-i\omega)}.
    \label{eq:SpectralDensitiesVelocities}
  \end{align}
\end{widetext}
Figure \ref{fig:Figure_Spectra} depicts a series of plots of the spectra given by Eq. \eqref{eq:SpectralDensitiesVelocities} that correspond to two points in Fig. \ref{fig:Fig_PerfectRectification}. For $c=1$ (Fig. \ref{fig:Figure_Spectra}(a) and (b)) there is no rectification, since the spectra match in the forward (a) and reversed (b) configurations. However, for $c=10$ ((Fig. \ref{fig:Figure_Spectra}(c) and (d))) the picture is very different: there is a good match between the spectra in the forward configuration whereas in the reversed configuration the spectra are less correlated, giving as a result higher rectification ($R \approx 0.8$). Figure \ref{fig:Figure_Spectra} only shows the elements (3,3) and (4,4) in the diagonal of $\mathbb{S}$ but the remaining elements, including off-diagonal ones, exhibit a similar behavior.
%
\section{Conclusions \label{sec:Conclusions}}
We have studied heat rectification in a model composed of two coupled harmonic oscillators connected to baths. This simple model allows analytical treatment but still has enough complexity to examine different ingredients that can produce rectification. 
Our results demonstrate in a simple but realistic system that harmonic systems can rectificate heat current if they have features which depend on the temperature  \cite{Pereira2017}. We implement this notion of temperature-dependent features by defining the baths exchange operation as an exchange of both temperatures and coupling parameters of the baths to the system. This kind of temperature-dependent features happens naturally in laser-cooled trapped ion set-ups.

We have also studied the phonon spectra of the system, comparing the match/mismatch of the phonon bands, to reach the conclusion that the band match/mismatch description for heat rectification is also valid for systems which are harmonic, as long as there are temperature-dependent features.
We hope this article sheds more light into the topic of heat rectification and that encourages more research regarding its physical implementation on chains of trapped ions.

\section{acknowledgements}
We thank Daniel Alonso for fruitful discussions and comments. This work was supported by the Basque Country Government (Grant No. IT986-16), by Grants PGC2018-101355-B-I00 (MCIU/AEI/FEDER,UE) and FIS2016-80681P, and by the Spanish MICINN and European Union (FEDER) (Grant No. FIS2017-82855-P). M.A.S. acknowledges support by the Basque Government predoctoral program (Grant No. PRE-2019-2-0234).

\bibliographystyle{apsrev4-1.bst}
\bibliography{Bibliography.bib}

\appendix

\section{Full set of steady-state equations for the components of $\mathbb{C}^{s.s}$  \label{AppStationaryStateEquations}}
Here we present the full set of equations for the covariance matrix elements in the steady state,
\begin{widetext}
  \begin{equation}
    \begin{split}
      \frac{2 k \expval{ p_2 q_1}^{s.s.} }{m_1 m_2}+\frac{2 \gamma _L \expval{ p_1^2}^{s.s.} }{m_1^3}&=\frac{2 D_L}{m_1^2},
      \\
      -\frac{2 k \expval{ p_2 q_1}^{s.s.} }{m_2^2}+\frac{2 \gamma _R \expval{ p_2^2}^{s.s.} }{m_2^3}&=\frac{2 D_R}{m_2^2},
      \\
      -\frac{\left(k_L+k\right) \expval{ q_1 q_2}^{s.s.} }{m_1}+\frac{k \expval{ q_2^2}^{s.s.} }{m_1}+\frac{\gamma _L \expval{ p_2 q_1}^{s.s.} }{m_1 m_2}+\frac{\expval{ p_1 p_2}^{s.s.} }{m_1 m_2}&=0,
      \\
      \frac{\left(k_L+k\right) \expval{ p_2 q_1}^{s.s.} }{m_1 m_2}-\frac{\left(k_R+k\right) \expval{ p_2 q_1}^{s.s.} }{m_2^2}+\frac{\gamma _L \expval{ p_1 p_2}^{s.s.} }{m_1^2 m_2}+\frac{\gamma _R \expval{ p_1 p_2}^{s.s.} }{m_1 m_2^2}&=0,
      \\
      -\frac{\left(k_L+k\right) \expval{ q_1^2}^{s.s.} }{m_1}+\frac{k \expval{ q_1 q_2}^{s.s.} }{m_1}+\frac{\expval{ p_1^2}^{s.s.} }{m_1^2}&=0,
      \\
      -\frac{\left(k_R+k\right) \expval{ q_2^2}^{s.s.} }{m_2}+\frac{k \expval{ q_1 q_2}^{s.s.} }{m_2}+\frac{\expval{ p_2^2}^{s.s.} }{m_2^2}&=0,
      \\
      -\frac{\left(k_R+k\right) \expval{ q_1 q_2}^{s.s.} }{m_2}+\frac{k \expval{ q_1^2}^{s.s.} }{m_2}-\frac{\gamma _R \expval{ p_2 q_1}^{s.s.} }{m_2^2}+\frac{\expval{ p_1 p_2}^{s.s.} }{m_1 m_2}&=0
    \end{split}
    \label{eq:SteadyStateEquationsModelB_Explicite}
  \end{equation}
\end{widetext}
\section{Complete expressions for the Spectral Density Matrix\label{AppSpecDenMat}}
In Section \ref{sec:lookingForR} we used the characteristic polynomial $P_{\mathbb{A}}(\lambda)$ of the dynamical matrix $\mathbb{A}$ for the calculation of the spectral density matrix. $P_{\mathbb{A}}(\lambda)$ is defined as
\begin{equation}
  \begin{split}
    \det(\mathbb{A}-\lambda) &= \lambda ^4 \\&+ \lambda ^3 \left(\frac{\gamma_L}{m_1}+\frac{\gamma_R}{m_2}\right) \\ &+ \lambda^2\frac{ (\gamma_L \gamma_R+m_2 (k+k_L)+m_1 (k+k_R))}{m_1 m_2}\\ &+ \lambda \frac{  (\gamma_R (k+k_L)+\gamma_L (k+k_R))}{m_1 m_2}\\ &+\frac{k (k_L+k_R)+k_L k_R}{m_1 m_2}.
  \end{split}
\end{equation}
We also used the polynomials $\mathbb{S}_L(\lambda)$ and $\mathbb{S}_R(\lambda)$, which are defined as $\mathbb{S}_L(\lambda)=\sum\limits_{n=0}^6 \lambda^n \mathbb{s}_{L,n}$ and $\mathbb{S}_R(\lambda)=\sum\limits_{n=0}^6 \lambda^n \mathbb{s}_{R,n}$. There are 14 different polynomial coefficients, which are $4\times 4$ matrices, which makes very cumbersome to include them in the main text. This is the full list of coefficients,
\begin{widetext}
  \small
  \begin{equation}
    \begin{aligned}
      \mathbb{s}_{L,0} &=
      \left(
      \begin{array}{cccc}
        (k+k_R)^2 & k (k+k_R) & 0 & 0 \\
        k (k+k_R) & k^2 & 0 & 0 \\
        0 & 0 & 0 & 0 \\
        0 & 0 & 0 & 0
      \end{array}
      \right),
      &
      \mathbb{s}_{R,0} & =
      \left(
      \begin{array}{cccc}
        k^2 & k (k+k_L) & 0 & 0 \\
        k (k+k_L) & (k+k_L)^2 & 0 & 0 \\
        0 & 0 & 0 & 0 \\
        0 & 0 & 0 & 0
      \end{array}
      \right),
      \\
      \mathbb{s}_{L,1} &= \left(
      \begin{array}{cccc}
        0 & k \gamma_R & -(k+k_R)^2 & -k (k+k_R) \\
        -k \gamma_R & 0 & -k (k+k_R) & -k^2 \\
        (k+k_R)^2 & k (k+k_R) & 0 & 0 \\
        k (k+k_R) & k^2 & 0 & 0
      \end{array}
      \right),
      &
      \mathbb{s}_{R,1} & = \left(
      \begin{array}{cccc}
        0 & -k \gamma_L & -k^2 & -k (k+k_L) \\
        k \gamma_L & 0 & -k (k+k_L) & -(k+k_L)^2 \\
        k^2 & k (k+k_L) & 0 & 0 \\
        k (k+k_L) & (k+k_L)^2 & 0 & 0
      \end{array}
      \right),
      \\
      \mathbb{s}_{L,2} &= \left(
      \begin{array}{cccc}
        2 (k+k_R) m_2-\gamma_R^2 & k m_2 & 0 & -k \gamma_R \\
        k m_2 & 0 & k \gamma_R & 0 \\
        0 & k \gamma_R & -(k+k_R)^2 & -k (k+k_R) \\
        -k \gamma_R & 0 & -k (k+k_R) & -k^2
      \end{array}
      \right),
      &
      \mathbb{s}_{R,2} & = \left(
      \begin{array}{cccc}
        0 & k m_1 & 0 & k \gamma_L \\
        k m_1 & 2 (k+k_L) m_1-\gamma_L^2 & -k \gamma_L & 0 \\
        0 & -k \gamma_L & -k^2 & -k (k+k_L) \\
        k \gamma_L & 0 & -k (k+k_L) & -(k+k_L)^2
      \end{array}
      \right),
      \\
      \mathbb{s}_{L,3} &= \left(
      \begin{array}{cccc}
        0 & 0 & \gamma_R^2-2 (k+k_R) m_2 & -k m_2 \\
        0 & 0 & -k m_2 & 0 \\
        2 (k+k_R) m_2-\gamma_R^2 & k m_2 & 0 & -k \gamma_R \\
        k m_2 & 0 & k \gamma_R & 0
      \end{array}
      \right),
      &
      \mathbb{s}_{R,3} & =\left(
      \begin{array}{cccc}
        0 & 0 & 0 & -k m_1 \\
        0 & 0 & -k m_1 & \gamma_L^2-2 (k+k_L) m_1 \\
        0 & k m_1 & 0 & k \gamma_L \\
        k m_1 & 2 (k+k_L) m_1-\gamma_L^2 & -k \gamma_L & 0
      \end{array}
      \right),
      \\
      \mathbb{s}_{L,4} &= \left(
      \begin{array}{cccc}
        m_2^2 & 0 & 0 & 0 \\
        0 & 0 & 0 & 0 \\
        0 & 0 & \gamma_R^2-2 (k+k_R) m_2 & -k m_2 \\
        0 & 0 & -k m_2 & 0
      \end{array}
      \right),
        &
        \mathbb{s}_{R,4} & = \left(
      \begin{array}{cccc}
        0 & 0 & 0 & 0 \\
        0 & m_1^2 & 0 & 0 \\
        0 & 0 & 0 & -k m_1 \\
        0 & 0 & -k m_1 & \gamma_L^2-2 (k+k_L) m_1
      \end{array}
      \right),
      \\
      \mathbb{s}_{L,5} &= \left(
      \begin{array}{cccc}
        0 & 0 & -m_2^2 & 0 \\
        0 & 0 & 0 & 0 \\
        m_2^2 & 0 & 0 & 0 \\
        0 & 0 & 0 & 0
      \end{array}
      \right),
      &
      \mathbb{s}_{R,5} & = \left(
      \begin{array}{cccc}
        0 & 0 & 0 & 0 \\
        0 & 0 & 0 & -m_1^2 \\
        0 & 0 & 0 & 0 \\
        0 & m_1^2 & 0 & 0
      \end{array}
      \right),
      \\
      \mathbb{s}_{L,6} &= \left(
      \begin{array}{cccc}
        0 & 0 & 0 & 0 \\
        0 & 0 & 0 & 0 \\
        0 & 0 & -m_2^2 & 0 \\
        0 & 0 & 0 & 0
      \end{array}
      \right),
      &
      \mathbb{s}_{R,6} & = \left(
      \begin{array}{cccc}
        0 & 0 & 0 & 0 \\
        0 & 0 & 0 & 0 \\
        0 & 0 & 0 & 0 \\
        0 & 0 & 0 & -m_1^2
      \end{array}
      \right)\,.
    \end{aligned}
  \end{equation}
\end{widetext}

\end{document}